\documentstyle[12pt]{article}
\newcommand{\be}{\begin{equation}}
\newcommand{\ee}{\end{equation}}
\newcommand{\ba}{\begin{eqnarray}}
\newcommand{\ea}{\end{eqnarray}}

\newcommand{\dsl}
  {\kern.06em\hbox{\raise.15ex\hbox{$/$}\kern-.56em\hbox{$\partial$}}}
\newcommand{\Dsl}{\not\!\! D}
\newcommand{\Psl}{\not\!\! P}
\newcommand{\eeq}{\end{equation}}
\newcommand{\eeqarr}{\end{eqnarray}}
\newcommand{\ZZ}{{\rm \kern 0.275em Z \kern -0.92em Z}\;}
\begin{document}
\title{Bogomol'nyi Bounds and the Supersymmetric Born-Infeld Theory}
\author{S.~Gonorazky\thanks{FOMEC-UNLP}\, ,
C.~N\'u\~nez\thanks{CONICET} \, , \\
F.A.~Schaposnik\thanks{Investigador CICBA} \, and
G.~Silva$^\dagger$
\\
{\normalsize\it
$^c$Departamento de F\'\i sica, Universidad Nacional de La Plata}\\
{\normalsize\it
C.C. 67, 1900 La Plata, Argentina}}
 
\date{\hfill}
\maketitle
\vspace{-3.5 in}

\hfill\vbox{
\hbox{~~}
\hbox{\it LA PLATA-Th 98/10}
}
\vspace{4.5 in}

\begin{abstract}
We study $N=2$ supersymmetric Born-Infeld-Higgs theory in $3$ dimensions
and derive Bogomol'nyi relations in its bosonic sector.
A peculiar coupling between the Higgs and
the gauge field  (with dynamics determined by the Born-Infeld action)
is forced by supersymmetry. The resulting equations coincide with those arising
in the Maxwell-Higgs model. Concerning
Bogomol'nyi bounds for the vortex energy, they
are derived from the $N=2$ supersymmetry algebra.

\end{abstract}


\bigskip

\newpage

\section{Introduction}

The supersymmetric extension of the
Born-Infeld theory \cite{B}-\cite{BI} was studied in
refs.~\cite{DP}-\cite{CF} by means of superspace techniques. Remarkably,
connections between Euler-Heisenberg effective Lagrangians
derived from certain supersymmetric theories and the Born-Infeld
Lagrangian were discovered \cite{DP},\cite{H}. More recently,
supersymmetric extensions of $10$-dimensional Born-Infeld
theory have been shown to play a central r\^ole
in the dynamics of D-branes \cite{Tse}-\cite{G}.

Closely related to the issue of supersymmetry completion
of the Born-Infeld theory, the  study of Bogomol'nyi relations
and BPS solutions in this theory is the main object of the present work.
To this end, we center the analysis in the study of a $N=2$
supersymmetric Born-Infeld theory in $d=3$ dimensions, which,
when coupled to a Higgs field, has a bosonic
sector which admits Bogomol'nyi equations \cite{NS1}-\cite{more}
\footnote{Originally,
Bogomol'nyi equations were discovered in
a $d=3$  model with a  Maxwell Lagrangian determining the
dynamics of the gauge field \cite{Bogo}- \cite{dVS}.
Already in \cite{dVS} the connection with supersymmetry is
signaled}. Interestingly
enough, the Born-Infeld BPS equations coincides with those
of the Maxwell theory and hence the exact vortex solutions
found in this last case \cite{dVS} also solve the more involved
Born-Infeld theory.

As it is well-known,
Bogomol'nyi relations can be found just by establishing
an inequality between  the energy and the topological charge
\cite{Bogo} or
by analysing the conditions under which a bosonic theory with
topological solutions can be extended to a $N=2$ supersymmetric
theory in which the central charge coincides with the
topological charge \cite{OW}. In this respect, it is very
enlighting to derive, via the Noether method, the explicit supersymmetric
algebra from which the origin and properties of BPS relations becomes
transparent. This was done for the Maxwell-Higgs model in
\cite{ed} and for the case of local supersymmetry in
\cite{bbs}-\cite{ed2}. In the present work we proceed to
a similar analysis with
the case of a supersymmetric Born-Infeld-Higgs  theory.

The plan of the paper is the following: in Section 2 we present
 the $N=1$ supersymmetric Born-Infeld theory in
$d=4$ dimensions giving
 an explicit formula for the fermionic Lagrangian
which will be necessary for constructing the SUSY charges. Then, in
Section 3 we proceed to a dimensional reduction to $d=3$ thus
obtaining a $N=2$ supersymmetric  Born-Infeld theory with a
bosonic sector obeying first order Bogomol'nyi equations. The
$N=2$ supersymmetric algebra is constructed in section 4 where
Bogomol'nyi bounds are discussed. Finally in section 5 we present
a discussion of our results. The explicit expressions for superfields
in components are detailed in an Appendix.

\section{Supersymmetric Born-Infeld theory}

In this section we shall start by writing the $N=1$
supersymmetric version of the
Born-Infeld (BI) theory in four dimensional
space-time. Then, by dimensional reduction, we shall obtain
in the next section a
$3$-dimensional $N=2$ supersymmetric Lagrangian which will
then be coupled
to a Higgs scalar.

The 4-dimensional Born-Infeld Lagrangian is
\begin{equation}
{L}_{{BI}}=\frac{\beta ^2}{e^2}\left(
1 - \sqrt{-\det \left(
g_{\mu \nu }+\frac 1\beta F_{\mu \nu }\right) } \right)
\label{1}
\end{equation}%
(The signature of the metric $g_{\mu\nu}$ is $(+,-,-,-)$).

Use of the identity
\be
\det \left( g_{\mu \nu }+\frac 1\beta F_{\mu \nu }\right)
= - 1-\frac 1{%
2\beta ^2}F^{\mu \nu}F_{\mu \nu}+
\frac 1{16\beta ^4}\left( F^{\mu \nu}\tilde F_{\mu \nu}
\right) ^2
\label{2}
\ee
allows to write (\ref{1}) in the form
\be
{L}_{{BI}}=\frac{\beta ^2}{e^2}\left(1 -
\sqrt{1+\frac 1{2\beta ^2}F^{\mu \nu}F_{\mu \nu}
-\frac 1{16\beta ^4}\left( F^{\mu\nu}\tilde F_{\mu \nu}\right) ^2}
\right)
\label{3}
\ee
Here,
$\tilde{F}_{\mu \nu} \equiv
\frac{1}{2}\varepsilon_{\mu \nu \rho \sigma}F^{\rho\sigma}$

In order to construct the SUSY extension of the Born-Infeld
Lagrangian, we shall follow \cite{DP}-\cite{CF}. Although
the complete derivation
of the bosonic part of the SUSY model has been presented
in these references (see also \cite{APS}-\cite{BG}) we shall
here give a detailed description of
the supersymmetric construction since, for
our purposes, knowledge of the explicit form
of certain fermion and fermion-boson terms is necessary.

We  start
by writing the BI Lagrangian (\ref{1}) in the form
\be
{ L}_{{BI}} =\frac{\beta ^2}{e^2}
\sum_{n=0}^\infty q_{n}\!
\left( \frac 1{2\beta ^2}F^{\mu \nu}F_{\mu \nu}-%
\frac 1{16\beta ^4}\left( F^{\mu \nu}\tilde F_{\mu \nu}
\right) ^2
\right)^{n+1}
\label{4}
\ee
where
\begin{eqnarray}
q_0 &=& - \frac{1}{2} \nonumber \\
q_n &=& \frac{\left( -1\right) ^{n+1}}{4^n}\frac{\left( 2n-1\right) !}
{\left(
n+1\right) !\left( n-1\right) !}   ~ ~ ~ ~ {\rm for} ~ ~ n \geq 1
\label{5}
\end{eqnarray}
Eq.(\ref{4}) can be rewritten as
\be
{ L}_{{BI}}=\sum_{n=1}^\infty q_{n-1}\sum_{j=0}^n
{n \choose j}
\left(  \frac 1{2\beta ^2}F^{\mu \nu}F_{\mu \nu}
\right) ^j\left( -\frac 1{16}
(\frac 1{\beta ^2}F^{\mu \nu}\tilde F_{\mu \nu} )^2\right) ^{n-j}
\label{8}
\ee
The basic ingredient for the supersymmetric extension of the
BI action is the curvature supermultiplet
\begin{equation}
W_\alpha =-\frac 14\bar D_{\dot\beta}\bar D^{\dot\beta} D_\alpha V
\label{9}
\end{equation}
where $V$ is the gauge vector superfield which in the Wess-Zumino
gauge reads
\be
V = - \theta \sigma^\mu \bar \theta A_\mu +
i \theta \theta \bar \theta \bar \lambda -
i \bar \theta \bar \theta \theta \lambda
+ \frac{1}{2} \theta  \theta  \bar \theta  \bar \theta D
\label{V}
\ee
Here $A_\mu$ is a vector field, $\lambda$ and
$\bar \lambda$ are two-component spinors
which can be combined to give  a four-component Majorana fermion
and $D$ is an auxiliary field. The covariant derivatives
 $D_\alpha$ and $\bar D_{\dot \alpha}$   act on chiral  variables
\begin{eqnarray}
y^\mu &=&x^\mu+i\theta \sigma ^\mu\bar \theta \nonumber\\
 y^{\mu \dagger} &=&x^\mu-i\theta \sigma ^\mu\bar \theta
\label{12}
\end{eqnarray}
where we use $\alpha, \beta, \ldots$ for spinor indices and
$\mu,\nu, \ldots $ for Lorentz indices. As usual,
$\sigma^\mu = (I,\vec \sigma)$ with $\sigma^i$ the Pauli matrices.
 Explicitly,
\begin{eqnarray}
D_\alpha &=& ~ \frac \partial {\partial \theta ^\alpha }+2i\left( \sigma
^\mu\bar
\theta \right) _\alpha \frac \partial {\partial y^\mu} \nonumber \\
\bar D_{\dot\alpha} & = &  - \frac{\partial}{\partial \bar\theta^{\dot\alpha}}
\label{10}
\end{eqnarray}
are the covariant derivatives acting on functions of $(y,\theta,\bar\theta)$
and
\ba
D_{\alpha} & = & \frac{\partial}{\partial \theta^\alpha} \nonumber \\
\bar D_{\dot \alpha } &=& - \frac
\partial {\partial \bar \theta ^{\dot \alpha }%
}-2i\left( \theta \sigma ^\mu\right) _{\dot \alpha }
\frac \partial {\partial
\bar y^\mu}
 \label{11}
\end{eqnarray}
are the corresponding covariant derivatives on functions of
$(y^\dagger,\theta,\bar\theta)$.

As it is well known, one can construct
the $N=1$ supersymmetric Maxwell Lagrangian
in terms of the chiral  superfield $W_\alpha$ and its hermitic conjugate
$\bar W_{\dot \alpha}$
by considering
\begin{equation}
L_0 = \frac{1}{4}
 \left[ \int d^2\theta W^2\left( y,\theta \right) +\int
d^2\bar \theta \bar W^2\left( y^{\dagger },\bar \theta \right) \right]
\label{12'}
\end{equation}
since the last $F$-components in $W^2$ and $\bar W^2$ contain
the terms $D^2-\frac{1}{2}\left( F^2\pm
\right.$ $\left. iF\tilde F\right)$. Now,
in order to get higher powers of $F^2$ and $F\tilde F$ necessary
to construct the BI action, it has been
shown  \cite{DP}
that one has to include powers of
superfields $X$ and $Y$ defined as
\be
X = \frac{1}{8} (D^2 W^2 + \bar D^2 \bar W^2)
\label{13}
\ee
\be
Y = -\frac{i}{16} (D^2 W^2 - \bar D^2 \bar W^2)
\label{14}
\ee
Note that, as shown in the Appendix, one can write
\be
X |_{\theta = \bar \theta = 0} =
-  (\frac{1}{\beta^2}
D^2 - \frac{1}{2 \beta^2} F^{\mu \nu} F_{\mu \nu}
-i \lambda \dsl \bar \lambda -i \bar \lambda \bar{\dsl}  \lambda )
\label{uno}
\ee
\be
Y|_{\theta = \bar \theta = 0} = \frac{1}{2} (\frac{1}{2 \beta^2}
F^{\mu \nu}\tilde F _{\mu \nu} +  \lambda \dsl \bar \lambda
- \bar \lambda \bar{\dsl}  \lambda)
\label{dos}
\ee
Hence,  lowest components of $X$ and $Y$ include the invariants $F^2$ and
$F \tilde F$.

Thus, we consider the following supersymmetric Lagrangian
whose bo\-so\-nic part, as we shall see, leads to the BI theory
\begin{equation}
{ L}_{{BI}}^{{SUSY}} =
\frac{\beta^2}{4e^2}\left[ \frac{1}{\beta^2}
\int d^2\theta W^2+
\frac{1}{\beta^2}\int d^2\bar
\theta \bar W^2\right] +
\sum_{r,s,t=0}^\infty a_{rst}\int d^4\theta \left(
W^2\bar W^2\right)^r X^s Y^t
\label{15}
\end{equation}
(here $a_{rst}$ are coefficients to be determined).
Concerning the last term in (\ref{15}), note that
 higher powers  of $F^2$  and $F \tilde F$ are
necessary for the construction of
the BI Lagrangian,
not only $W^2$ and $\bar W^2$  should be considered
but also products and powers of these chiral superfields
have to be introduced. Now, since
the highest component of $W^2 \bar W^2$  takes the form
\be
W^2 \bar W^2\vert_{\theta \theta \bar \theta \bar \theta} =
\theta \theta \bar \theta \bar \theta \left(
(D^2 - \frac{1}{2} F_{\mu \nu}F^{\mu \nu})^2 +
(\frac{1}{2}\tilde F_{\mu \nu}F^{\mu \nu})^2 \right)
\label{termino}
\ee
 one can see that  powers of this
general superfield  can be used to reproduce the
expansion in (\ref{8}).

As stated above, coefficients $a_{rst}$ should be chosen so as to get
the BI Lagrangian from the bosonic part of (\ref{15}). Since
the product
$X^{s} Y^t$ contains derivatives of $F$ and $\tilde F$
in its $D$-component, $a_{0st}$ must vanish
in order to eliminate  unwelcome purely bosonic terms.
Moreover, one can see that the simplest choice leading to the
BI action is to take $a_{rst} = 0$ for $r>1$. Then, in superfield
notation, the SUSY BI Lagrangian that we shall consider is:
\begin{equation}
{L}_{{BI}}^{{SUSY}} =
\frac{1}{4e^2}\left[ \int d^2\theta W^2+\int d^2\bar
\theta \bar W^2\right] +
\sum_{s,t=0}^\infty a_{1st}\int d^4\theta
W^2\bar W^2   X^s Y^t
\label{16}
\end{equation}
One  can see that this Lagrangian coincides with those proposed in
\cite{DP}-\cite{CF}.

It remains to determine coefficients $a_{1st}$ so
that the bosonic sector of the theory
does coincide with the BI Lagrangian.
We then concentrate in the purely bosonic terms
of ${L}_{{BI}}^{{SUSY}}$ and this we do by putting fermions
to zero. At this stage, we shall impose
\begin{eqnarray}
\left. {L}_{{BI}}^{{SUSY}}\right| _{%
{BOS}}
& \equiv & {\tilde L}_{{BI}}
 \nonumber\\
& =& \frac{\beta ^2}{e^2}\left(1 -
\sqrt{1+\frac 1{2\beta ^2}F^{\mu \nu}F_{\mu \nu}
-\frac 1{16\beta ^4}\left( F^{\mu\nu}\tilde F_{\mu \nu}\right) ^2 -
\frac{1}{\beta^2}D^2}
\right)\nonumber\\
\label{quien}
\end{eqnarray}
Note that when the equation of
motion for the $D$ field (which in the present case gives $D=0$) is used,
the bosonic part of the supersymmetric Lagrangian coincides
with the BI Lagrangian, i.e. ${\tilde L}_{BI}[D=0] = L_{BI}$.

Coefficients $a_{1st}$ can be now computed by imposing
identity (\ref{quien}). From the  bosonic components of $W^2 \bar W^2$
(given in the Appendix) one  finds a recurrence relation which connects
the $a's$ coefficients in SUSY BI Lagrangian expansion
with coefficients $q's$ (eq.(\ref{5})) in the
expansion of the BI Lagrangian,
\begin{eqnarray}
& & a_{100} = \frac{1}{8\beta^2} \nonumber\\
& & a_{1 ~ n - 2m ~ 2m}  =  \frac{(-1)^m}{\beta^{2n + 4}}
 \sum_{j=0}^m 4^{m-j}{{n + 2 - j} \choose j}
q_{n+1-j} \nonumber\\
& & a_{1 ~ n  ~ 2m + 1}  =  0
\label{21}
\end{eqnarray}
With the knowledge of coefficients $a's$, the supersymmetric Born-Infeld
Lagrangian can be explicitly written in the form
\be
{L}_{{BI}}^{{SUSY}} = \tilde L_{BI} + L_{fer} + L_{fb}
\label{22}
\ee
where $L_{fer}$ contains self-interacting fermion terms while
$L_{fb}$ includes kinetic fermion  and crossed
boson-fermion terms.
Both $L_{fer}$ and  $L_{fb}$
 can be calculated as
expansions in increasing powers of fermionic and bosonic fields.
For our purposes, namely the discussion of Bogomol'nyi relations
 through the supersymmetry algebra, only certain terms,
quadratic in fermion fields, will be necessary.

In fact, as will become clear
 in  section (IV) , only quadratic terms of the form
$ \lambda \partial_\mu \bar \lambda$ and
$ \bar \lambda \partial_\mu  \lambda$ will give a contribution
to the current algebra (higher order terms vanish when fermions
are put to zero). Then, we shall only give the explicit form
of those terms in $L_{fb}$
which will be necessary in what follows (terms in $L_{fer}$
do not contribute to the SUSY algebra).
Denoting $L_{fb}[ \lambda,  \partial \bar\lambda]$ the
sum of relevant terms
(other terms of equal or higher order in
fermionic fields  can be calculated straightforwardly),
we have
\be
L_{fb} = L_{fb}^I[ \lambda,  \partial \bar \lambda]
+ L_{fb}^{II}[ \bar \lambda, \partial\lambda] + ~ {\rm other ~~ terms}
\label{40}
\ee
\begin{eqnarray}
& &  L_{fb}^I\left[ \lambda ,\partial \bar \lambda \right]  =
-\frac{i}{2}
\lambda
\dsl \bar \lambda
-i\sum_{s,t=0}^\infty a_{1st}\lambda \sigma ^\nu \partial _\mu \bar
\lambda (X_{{BOS}})^{s-1}(Y_{{BOS}})^{t-1}
\nonumber \\
& & \left(
-2iX_{{BOS}}Y_{{BOS}}
+A^{*}(isY_{{BOS}}+\frac t2X_{{BOS}})
\right)
\left( A\delta _\nu ^\mu +\frac 12\Omega ^{*\mu \rho }\Omega _{\rho
\nu }\right)
\label{222}
\end{eqnarray}
\be
L_{fb}^{II}[ \bar \lambda, \partial\lambda] =
L_{fb}^I\left[ \lambda ,\partial \bar \lambda \right]^\dagger
\label{qqq}
\ee

Here A and $\Omega ^{*\mu \rho }\Omega _{\rho \nu }$, calculated
in the Appendix, are given by
\be
A=D^2-\frac 12F^{\mu \nu }F_{\mu \nu }-\frac i2F^{\mu \nu }\tilde F_{\mu \nu
}
\ee
\be
\Omega ^{*\nu \rho }\Omega _{\rho \mu }=\left( D^2+\frac 12F_{\alpha \beta
}F^{\alpha \beta }\right) \delta _\mu ^\nu -2D\eta ^{\nu \rho }\tilde
F_{\rho \mu }+2F^{\nu \rho }F_{\rho \mu }
\ee

We end this section by rewriting $L_{{BI}}^{{SUSY}}$ defined in
eq.(\ref{22}), which was worked out in terms of the two component fermions
$\lambda
$ and $\bar \lambda ,$ using a four component fermion $\Lambda ,$
\be
\Lambda =\left(
\begin{array}{c}
\lambda _\alpha  \\
\bar \lambda ^{\dot \alpha }
\end{array}
\right)
\label{c88}
\ee
Concerning 4 dimensional Dirac matrices $\Gamma ^\mu ,$ we use
\be
\Gamma ^\mu =\left(
\begin{array}{cc}
0 & \sigma ^\mu  \\
\bar \sigma ^\mu  & 0
\end{array}
\right)
\ee
Then, instead of eq.(\ref{40}), we have for $L_{fb}$
\begin{eqnarray}
L_{fb} &=&
-\frac{i}{2}
\bar \Lambda \dsl \Lambda  +
\sum_{s,t=0}^\infty \beta ^
{2 \left( s+2t+1 \right)}
a_{1st} X_{{BOS}} ^ {s-1} Y_ {{BOS}} ^ {2t-1}
\nonumber\\
& &
\left(
 i\bar \Lambda \dsl \Lambda Y_{{BOS}}
\left[
s(X_{{BOS}}^2+4Y_{{BOS}}^2)-X_{{BOS}}
\left( Z_{{BOS}}- 2X_{{BOS}}\right)
 \right]
\right.
\nonumber \\
& &
\left.
+2i\bar \Lambda \Gamma ^\mu \partial ^\nu \Lambda (D\tilde
F_{\nu\mu }-F_{\nu \rho }F_{\;\mu }^\rho )
\left
[ X_{{BOS}}Y_{{BOS}}+
2
\left
( 2sY_{{BOS}}^2+tX_{{BOS}}^2
\right)
\right]
 \right.
\nonumber \\
& &
+\bar \Lambda \Gamma ^5 \dsl \Lambda X_{BOS}
\left[
t(X_{BOS}^2+4Y_{BOS}^2)+
4Y_{{BOS}}^2+
 Y_{{BOS}}Z_{{BOS}}(s-2t)\right]
\nonumber \\
& &
\left. -2\bar \Lambda \Gamma ^5\Gamma ^\mu \partial ^\nu \Lambda (D\tilde F%
_{\nu \mu }-F_{\nu \rho }F_{\;\mu }^\rho )X_{{BOS}}Y_{{BOS}%
}(s-2t)\right)  \label{ui}
\end{eqnarray}
where
\begin{eqnarray}
X_{{BOS}} &=&-D^2+\frac 12F_{\mu \nu }F^{\mu \nu }
\nonumber\\
Y_{{BOS}} &=&\frac 14F_{\mu \nu }\tilde F^{\mu \nu }
\nonumber\\
Z_{{BOS}} &=&D^2+\frac 12F_{\mu \nu }F^{\mu \nu }
\end{eqnarray}
\noindent With this, the complete supersymmetric Born-Infeld Lagrangian
(\ref{22})
\[
{L}_{{BI}}^{{SUSY}} = \tilde L_{BI} + L_{fer} + L_{fb}
\]
is invariant under the following $N=1$ supersymmetry transformations
\begin{eqnarray}
& & \delta A_\mu  =  - i \bar \epsilon \Gamma_\mu \Lambda  \;\;\;\;\;
\;\;\;\;\;
\delta \Lambda = i (-\Sigma^{\mu \nu}F_{\mu \nu} + \Gamma^5 D)\epsilon
\nonumber\\
& & \delta D  =  i \bar \epsilon \Gamma^5 \dsl \Lambda
\label{SN1}
\end{eqnarray}
where
$\Sigma_{\mu \nu} = \frac{i}{4}[\Gamma_\mu,\Gamma_\nu]$
and $\Gamma^5 = i \Gamma^1\Gamma^2\Gamma^3\Gamma^0$.

\section{The Supersymmetric Born-Infeld Higgs \hfill\break
model
and Bogomol'nyi equations}

In the previous section we have constructed the $d=4$, $N=1$
supersymmetric Born-Infeld theory.
Now, since we are seeking for Bogomol'nyi relations
for a spontaneously broken gauge theory, one
has to consider, in addition to the Lagrangian already derived,
a SUSY Higgs Lagrangian. This can be done by considering
a chiral supermultiplet $\Phi$
coupled to the vector superfield (\ref{V}) in the usual gauge invariant way.
Also, one adds a Fayet-Iliopoulos term to break the gauge symmetry. We shall
not give the details here but directly give the resulting SUSY Higgs
Lagrangian.

The first part of  this section is devoted to a
 dimensional reduction to $d=3$ space-time
thus obtaining a $N=2$ supersymmetric theory. It is in this model
that Bogomol'nyi relations for vortices arise. Then, we shall discuss
Bogomol'nyi equations.

Now, as it is well known,  enlargement of supersymmetry from $N=1$ to $N=2$
leads to a bosonic sector
obeying first order Bogomol'nyi equations
\cite{LLW}-\cite{ed}. These equations are obtained at the end of this
section while the supersymmetry algebra (leading to Bogomol'nyi bounds)
is discussed in the next one.

The dimensional reduction proceeds as follows. We take $A_3 = N$ and
$N$ will play the role of an additional scalar field in the
$3$-dimensional model. Concerning the fermion $\Lambda$ defined in
eq.(\ref{c88}), its components can be accomodated
into a couple of
two-component $3$-dimensional Majorana fermion $\chi$ and $\rho$
Four dimensional $\Gamma$ matrices are related to
$2 \times 2$ Dirac matrices in three
dimensions $\gamma^i$ ($i = 0,1,2$) as follows
\[ \Gamma^i = \gamma^i\otimes\tau_3 \;\;\; , \;\;\;
\Gamma^3 = 1\otimes i\tau_2  \;\;\; , \;\;\;
\Gamma^5 = 1\otimes \tau_1  \]
\be
\Sigma^{ij} = \sigma^{ij}\otimes 1 \;\;\; , \;\;\;
\Sigma^{i3} = -\Sigma^{3i} = \gamma^i\otimes\tau_1
\label{gammas}
\ee
where $\sigma^{ij} = 1/2 [\gamma^i,\gamma^j]$.

With this, the dimensionally reduced $N=2$, $d=3$
action takes the form
\be
S^{(3)} = S_{bos}^{(3)} + S_{fb}^{(3)} + S_{fer}^{(3)}
\label{c9}
\ee
Here
\begin{eqnarray}
 S_{bos}^{(3)}   &=& -\frac{\beta^2}{e^2} \int d^3x
\nonumber\\
& &
 \left( \sqrt{1 - \frac{1}{\beta^2}D^2   + \frac{1}{2\beta^2} F^{ij}F_{ij}
-\frac{1}{\beta^2}\partial_iN\partial^iN  -
\frac{1}{\beta^4}(\varepsilon_{ijk}F^{ij}\partial^k N)^2}
- 1\right)\nonumber\\
\label{c10}
\end{eqnarray}
Concerning $S_{fb}^{(3)}$, it can be written in the form
\begin{eqnarray}
S_{fb} &=&-\frac i2\int d^3x\bar \Sigma \dsl\Sigma
+i\sum_{s,t=0}^\infty \frac{a_{1st}}{\beta ^{2s+4t+2}}
\left(
X^{\left(3\right) }
\right)
 ^{s-1}
\left(
Y^{
\left( 3\right)
 }
\right) ^{2t-1}
\nonumber \\
&&
\left \{
\bar \Sigma \dsl \Sigma
\left[
s
\left(
 \left( X^{\left(3\right) }\right) ^2+
\left( Y^{\left( 3\right) }\right) ^2
\right)
 Y^{\left(3\right) }+
\left(
 X^{\left( 3\right)
 } - 2Z^ {
\left( 3\right) }
\right)
X^{
\left( 3\right)
 } Y^{
\left( 3\right)
 }
+
\right.
\right.\nonumber
\\
&&
\left.
 \frac{1}{2} t X^{
\left( 3 \right)}
\left(
 \left( X^{\left(3\right) }
\right) ^2+
\left(
Y^{\left( 3\right) }
\right) ^2
\right)
+\left(
Y^{\left( 3\right) }
\right)
 ^2 X^{\left( 3\right) }
\right]
-
\bar \Sigma \gamma ^i\partial ^j\Sigma
\nonumber\\
& &
\left(
 X^{
\left( 3\right)
}
\eta _{ij}+2F_{ik}F_{\;j}^k
\right)
\left[
X^{\left( 3\right) }
Y^{\left(3\right) }
\right.
( F_{ik}F_{\;j}^k  +
\left. \right.
\left. \right.
  \frac{s}{2}-t) \nonumber\\
& &
\left.
+ \left(
s\left( Y^{\left( 3\right) }\right) ^2
 + t\left( X^{\left( 3\right) }\right) ^2
\right)
\right]
 \nonumber
\\
&&
+ \left.
\bar \Sigma \partial ^j\Sigma D\tilde F_j
\left[
\left( Y^{\left(3\right) }\right) ^2
+ 2t\left( X^{\left( 3\right) }\right) ^2-
\left(
1 + 2s - 4t \right) X^{\left( 3\right) }
Y^{\left( 3\right) }\right] \right\}
\label{ff}
\end{eqnarray}

where

\begin{eqnarray}
X^{\left( 3\right) } = \frac 12 F_{ij}F^{ij}  - D^2  -
\left( \partial _iN\right)^2 \\
Y^{\left( 3\right) } = \tilde F^i\partial _iN \\
Z^{\left( 3\right) } = D^2+\frac 12 F_{ij}F^{ij}
\label{ff1}
\end{eqnarray}
and $\Sigma$ is a Dirac fermion constructed from the two
Majorana fermions $\chi$ and $\rho$,
\be
\Sigma = \chi + i\rho
\label{c12}
\ee
Finally, $S_{fer}^{(3)}$ is the dimensionally reduced purelly
fermionic action whose explicit form is irrelevant
for our main purpose, namely the evaluation of the supersymmetry algebra.

As announced, we shall add  a $N=2$, $d=3$
Higgs action $S_H^{(3)}$
for the Higgs field whic takes the form \cite{ed}
\begin{eqnarray}
S_H^{(3)} & = &
\int d^3x
\left(\frac{1}{2}|D_i\phi|^2 +
 \frac{i}{2} \bar \psi \Dsl \psi
+ \frac{1}{2} |F|^2 +
\frac{i}{2}(\bar\psi \Sigma \phi -  \bar \Sigma \psi \phi^*)
+  \right.
\nonumber\\
& &  \frac{D}{2}(|\phi|^2 - \xi^2) +
\left. \frac{1}{2} N(F \phi^* + F^*\phi -
\bar \psi \psi)
\right)
\label{c3}
\end{eqnarray}
$\phi$ is a complex charged scalar, $\psi$ a Dirac
spinor, $N$ a real scalar  and $F$ a complex auxiliary field.
The covariant derivative
\be
D_i = \partial_i + i A_i
\label{c4}
\ee
Using the equation of motion for the auxiliary field $F$, action
(\ref{c3}) reads
\begin{eqnarray}
S_H^{(3)} & = & \int d^3x
\left(
\frac{1}{2}|D_i\phi|^2 +
 \frac{i}{2} \bar \psi \Dsl \psi +
\frac{i}{2}(\bar\psi \Sigma \phi -  \bar \Sigma \psi \phi^*)
+ \frac{D}{2}(|\phi|^2 - \xi^2)  -  \right.
\nonumber\\
& &
\left.
\frac{1}{2}N^2 |\phi|^2 - \frac{1}{2} N \bar \psi \psi
\right)
\label{F}
\end{eqnarray}

The complete $N=2$, $d=3$ supersymetric Born-Infeld-Higgs
action is then given by
\be
S^{(3)}_{SUSY} = S^{(3)}_{bos}  + S^{(3)}_{fb} +
S^{(3)}_{fer} + S_H^{(3)}
\label{c6}
\ee
where the different actions have been defined through
eqs.(\ref{c9})-(\ref{ff}) and (\ref{F}). Dimensions of
parameters and fields in units of
mass are: $[\beta] = m^2$, $[e] = m^\frac 12$, $[\xi] = m^\frac 12$,
$[(A_\mu,N,\Sigma,D)] = (m,m,m^\frac32,m^2)$ and
$[(\phi,\psi,F))] = (m^\frac 12, m,m^\frac 32)$.

Action (\ref{c6}) remains invariant under the following $N=2$
supersymmetry
transformations (with Dirac fermion parameter $\epsilon$)
 \be
\begin{array}{lll}
\delta \phi =  \bar \epsilon \psi  &  ~ ~
\delta \psi = -(i \Dsl \phi + N \phi)\epsilon & ~ ~
\delta N = i \bar \epsilon \Sigma + {\rm h.c.}
\nonumber \\
\delta A_i =  \bar \epsilon \gamma_i \Sigma  &  ~ ~
\delta \Sigma = (\frac{1}{2}\varepsilon_{ijk}F^{ij}\gamma^k + D
                 + i\dsl N) \epsilon &  ~ ~
\delta D = \frac{1}{2} \bar \epsilon  \dsl \Sigma - {\rm h.c.}
\end{array}
\label{susyy}
\ee
Since we have already used the equation of motion of the $F$ field,
eqs.(\ref{susyy}) correspond to an on-shell invariance. In
order to have an off-shell invariance one
just has to use (\ref{c3}) instead of (\ref{F}) and supplement
(\ref{susyy}) with the transformation law for $F$
\be
\delta F = i \bar \epsilon \Dsl \psi +
( i \bar \epsilon\Sigma \phi + {\rm h.c.} )
\label{sss}
\ee

The connection between supersymmetry and Bogomol'nyi equations is by
now well-known. In the ``normal'' Maxwell-Higgs theory,
imposing  the  supersymmetry variation of the gaugino to be zero gives one
of the Bogomol'nyi equations (that for the gauge curvature) while
the vanishing of the Higgsino supersymmetry variation leads to the
second Bogomol'nyi equation, the one for the Higgs field. The same
happens in the present case. Indeed, suppose we want to obtain the
Bogomol'nyi equations for the bosonic theory defined by action
(\ref{c6}) with all fermion fields put to zero. If we use the
equation of motion for the auxiliary field $D$,
\begin{eqnarray}
D & = & - \frac{e^2}{2}
\frac{\phi^2 - \xi^2}{\sqrt{1 + \frac{e^4}{4\beta^2}
(\phi^2 - \xi^2)^2}} \times\nonumber\\
& &
\sqrt{1 +
\frac{1}{2\beta^2} F^{ij}F_{ij} -
\frac{1}{\beta^2} \partial_iN \partial^iN-
\frac{1}{\beta^4} (\varepsilon_{ijk}F^{ij}\partial^k N)^2
}
\label{lab}
\end{eqnarray}
the bosonic action becomes
\begin{eqnarray}
& & S \equiv S^{(3)}_{bos} + S^{(3)}_H =  \frac{\beta^2}{e^2} -\nonumber\\
& & \frac{\beta^2}{e^2}
\int d^3x
\sqrt{
\left(
1 +
\frac{1}{2\beta^2} F^{ij}F_{ij} -
\frac{1}{\beta^2} \partial_iN \partial^iN -
\frac{1}{\beta^4} (\varepsilon_{ijk}F^{ij}\partial^k N)^2
\right)
V[\phi]}
 \nonumber\\
\label{ac}
\end{eqnarray}
where $V[\phi]$ is the resulting symmetry breaking potential
\be
V[\phi] =
1 + \frac{e^4}{4\beta^2}(\phi^2 - \xi^2)^2
\label{sb}
\ee
It is interesting to note that in our treatment, the
symmetry breaking potential appears as a multiplicative
factor inside the BI square root as a result of searching
the supersymmetric extension of the bosonic theory. That is,
{$N = 2$ \it supersymmetry forces this functional
form for the action} (the same happens if one remains in
$d=4$ dimensions with the $N=1$ theory). In ref. \cite{NS1}
this functional form was selected from the infinitely many possibilities
of adding to the BI theory
a Higgs field and its symmetry breaking potential just by trying to
obtain the usual (i.e. Maxwell+Higgs) Bogomol'nyi
equations. Thus, supersymmetry explains the rationale of the choice
associated with the Born-Infeld-Higgs model.

Let us now   write, exploiting supersymmetry,
 the first order
equations for the Born-Infeld-Higgs
theory which are the analogous to the Bogomol'nyi equations for the
``normal'' Nielsen-Olesen model. To this end we consider the static case with
$A_0 = N = 0$.

{}From the supersymmetry point of view, Bogomol'nyi equations
follow from the analysis of the gaugino and Higgsino supersymmetry
variations. More precisely, one decomposes the (Dirac) parameter of
the supersymmetry transformation into its chiral components
$\epsilon_\pm$. Then, by imposing the vanishing of half of
the supersymmetry variations, say those generated by $\epsilon_+$
($\epsilon_-$) one gets the Bogomol'nyi equations in a
soliton (anti-soliton) background. The other half supersymmetry
is broken.
In the present case this amounts, for a vortex with
\underline{positive} magnetic flux, to the conditions
\begin{eqnarray}
\delta_{\epsilon_+} \Sigma = 0 & \to & \frac{1}{2}\varepsilon_{0ij}F^{ij}
=
-D
\label{bo1}\\
\delta_{\epsilon_+} \psi = 0 & \to & D_1\phi = i D_2 \phi
\label{bogo}
\end{eqnarray}
\begin{eqnarray}
\delta_{\epsilon_-} \Sigma & \ne & 0 \nonumber\\
\delta_{\epsilon_-} \psi & \ne & 0
\label{bogos}
\end{eqnarray}
Using the explicit expresion given by (\ref{lab}),
we can rewrite (\ref{bo1}) in the form
\be
\frac{1}{2}\varepsilon_{0ij}F^{ij}
=
 \frac{e^2}{2}
\frac{\phi^2 - \xi^2}{\sqrt{1 + \frac{e^4}{4\beta^2}
(\phi^2 - \xi^2)^2}}
\sqrt{1 +
\frac{1}{2\beta^2} F^{ij}F_{ij}
}
\label{epa}
\ee
{}From this equations, we obtain a simple expression for the
magnetic field  which
in fact coincides with that corresponding to the ``normal''
(i.e. with Maxwell dynamics) Bogomol'nyi equation
\be
B \equiv (1/2)\varepsilon_{0ij}F^{ij} =  \frac{e}{2} (|\phi|^2 - \xi^2)
\label{B}
\ee
This equation, together with  (\ref{bogo}) are the Bogomol'nyi
equations for the Born-Infeld-Higgs system. They coincide with
those arising in the Maxwell-Higgs system, i.e., the original
Bogomol'nyi equations \cite{Bogo}-\cite{dVS} and hence
they have the same exact solutions originally found  in \cite{dVS}

\section{SUSY algebra}
Given the $3$ dimensional model defined by action (\ref{c9}), one can
easily construct the associated conserved supercurrent and from it
the supercharge commutators. The
corresponding supercharges $\bar Q$ and $Q$ can be written in the
following form
\be
\bar Q \epsilon \equiv \int d^2x
( \frac{\partial L}{\partial(\partial_0\Sigma)} \delta\Sigma
+
\frac{\partial L}{\partial(\partial_0\psi)} \delta \psi )
\label{def}
\ee
\be
Q \equiv \gamma^0  \bar Q^\dagger
\label{defi}
\ee
After some work one gets
\begin{eqnarray}
\bar Q \!\!\! & = & \!\!\!\frac{i}{2e^2}
\int d^2x
\Sigma^\dagger
\left(
1 + 2\sum_0^\infty a_{1s0} \beta^{-2(s+1)}(B^2 - D^2)^s
((2s + 3)D^2 - \right.\nonumber\\
& &  \left. \vphantom{\sum_0^\infty}
 B^2  - 2(s+1) \gamma^0 BD )
\right) ( \gamma^0 B + D) +
 \frac{i}{2} \int d^2x \psi^\dagger {\Dsl \phi}
\label{qra}
\end{eqnarray}
\begin{eqnarray}
Q\!\!\! & = & \!\!\!-\frac{i}{2e^2}
\int d^2x
(B + \gamma^0 D)
\left(
1 + 2\sum_0^\infty a_{1s0} \beta^{-2(s+1)}(B^2 - D^2)^s
 \right.\nonumber \\
& &
\left. \vphantom{\sum_0^\infty}
((2s + 3)D^2 -
B^2  - 2(s+1) \gamma^0 BD )
\right) \Sigma -
\frac{i}{2} \int d^2x
({\Dsl \phi})^*
\psi
 \label{qraf}
\end{eqnarray}
As in the previous section, we have considered Nielsen-Olesen
vortices by putting $A_0 = N = 0$ after differentiation
(We also restrict ourselves to the static
case). More important, we have only included terms linear in the fermionic
fields, this because we are interested in extracting, from the
SUSY charge algebra, just the pure bosonic term from which the
(bosonic) Bogomol'nyi equations will be derived.
That is why, after computing the algebra, all fermion
fields should be put to zero (Non-linear fermionic terms
in the charges necessarily give fermionic contributions to the
algebra which vanish when fermions are put to zero).

Our purpose is to compute the Born-Infeld SUSY charge algebra and, from it,
to explicitly obtain the Bogomol'nyi bounds in terms of energy
and central charge.
Since the expansion of the
Born-Infeld Lagrangian in powers of $1/\beta^2$ leads to Maxwell,
Euler-Heisenberg, ... Lagrangians, it will be instructive to
 show how the algebra reproduces, in a
$1/\beta^2$ expansion, the  corresponding Maxwell, Euler-Heisenberg,
... SUSY algebra and then present the arguments leading to the complete
result.
Indeed, to zero order in $1/\beta^2$ one gets for the
SUSY charges, which we denote to this order as $\bar Q^{(0)}$ and
$ Q^{(0)}$,
\be
\bar Q^{(0)} = \frac{i}{2e^2} \int d^2x \Sigma^\dagger (\gamma^0
B + D)+ \frac{i}{2} \int d^2x \psi^\dagger \Dsl \phi
\label{mm}
\ee
With this and $Q^{(0)}$
which can be computed from  eq.(\ref{defi}), one can compute
the SUSY algebra which takes the form
\be
\{ Q^{(0)},\bar Q^{(0)}\} = \Psl + Z
\label{z}
\ee
with $P_\mu$  the $4$-momentum and  $Z$ the central charge. Using
the explicit forms for $\bar Q^{(0)}$ and $ Q^{(0)} $ obtained above
one can compute the Poisson bracket corresponding to the l.h.s. in
(\ref{z}) and, comparing with the r.h.s. in this last equation
one can identify
\be
P^0 = {\rm tr} (\gamma^0 \{ Q^{(0)},\bar Q^{(0)}\})
= \int d^2x \left(\frac{1}{2e^2} B^2 +  \frac{1}{2}| D_i\phi|^2 +
\frac{e^2}{8}(\phi^2 - \xi^2)^2
\right)
\label{qqs}
\ee
\be
Z = {\rm tr} (\{ Q^{(0)},\bar Q^{(0)}\})
= \int d^2x \left(
B(\phi^2 - \xi^2) + \varepsilon_{ij}D_i\phi (D_j\phi)^*
\right) = \xi^2 \oint A_\mu  dx^\mu
\label{uu}
\ee
As it is well-known \cite{OW}, hermiticity of anticommutator (\ref{z}) leads
to a Bogomol'nyi bound which in the present case corresponds
to the Maxwell-Higgs theory,
\be
P^0 = E \ge |Z|
\label{bm}
\ee
or
\be
E \ge \xi^2 2\pi n
\label{e}
\ee
where $n$ is the number of flux lines measured by $Z$ \cite{ed}.

We now consider the next order in the $1/\beta^2$ expansion, namely
the Euler-Heisenberg theory. In that case, instead of (\ref{mm}) we have
\be
\bar Q^{(1)} = \bar Q^{(0)} + \frac{i}{2e^2} \int d^2x \Sigma^\dagger
\left( \frac{1}{4\beta^2}(3D^2 - B^2) - \frac{1}{2\beta^2}\gamma^0 B D
\right)
(\gamma^0 B + D)
\label{eee}
\ee
The SUSY charges anticommutator leads in this case to
\begin{eqnarray}
P^0 = E & = &
\int d^2x \left(\frac{1}{2e^2} B^2 +  \frac{1}{2}| D_i\phi|^2 +
\frac{e^2}{8}(\phi^2 - \xi^2)^2 -
\right.
\nonumber \\
& &
 \left. \frac{1}{8\beta^2e^2}\left(B^2 -  \frac{e^4}{4}(\phi^2 -
\xi^2)^2\right)^2
\right)
\label{qqse}
\end{eqnarray}
$Z$ is still given by (\ref{uu}) and eq.(\ref{e}) also holds in this case.

The next order leads to the following results
\begin{eqnarray}
\bar Q^{(2)} &=& \bar Q^{(1)} +  \frac{i}{2e^2\beta^4}
\int d^2x \Sigma^\dagger
\frac{1}{2}(B^2 - D^2)\left((5D^2 - B^2) -
\frac{1}{4}\gamma^0 B D
\right) \nonumber\\
& & (\gamma^0 B + D)
\label{doss}
\end{eqnarray}
\begin{eqnarray}
P^0 = E\!\!\! & = &\!\!\! \int d^2x\frac{1}{2e^2}
\left
( B^2 +
\frac{e^4}{4}(\phi^2 - \xi^2)^2
\right)
\left(
 \frac{1}{2}| D_i\phi|^2
-\frac{1}{4\beta^2}\left(B^2 -\right.\right.\nonumber \\
& & \left. \left. \frac{e^4}{4}(\phi^2 - \xi^2)^2\right) +
\frac{1}{16 \beta^4 e^2}
\left(B^2 -\frac{e^4}{4}(\phi^2 - \xi^2)^2\right)
\right)
\nonumber \\
& & \left(B^2 +\frac{e^4}{4}(\phi^2 - \xi^2)^2\right)^2
\label{qqse3}
\end{eqnarray}
Again, $Z$ is given by eq.(\ref{uu}), the same expression as in the Maxwell
and Euler-Heisenberg case. In fact, this coincidence is not accidental and
one can understand it as follows. If one were to obtain the Bogomol'nyi
bound not from supersymmetry but as originally done by Bogomol'nyi, one
should look at the purely bosonic Born-Infeld-Higgs theory and write
the energy as a sum of squares plus a surface term. This surface term
is responsible for the appearence of the topological charge
as the bound for the energy. Now, the surface term is not modified
by the fact that one deals with a BI and not a Maxwell gauge
field Lagrangian. Moreover, the Bogomol'nyi equations do coincide
for these two theories. Viewed from the supersymmetry side, the
bound for the energy is provided by the central charge which again
does not depend on the form of gauge field kinetic energy term.

Coming back to the complete SUSY algebra, let us write  the charge $\bar Q$
given by
eq.(\ref{qra})  in the form:
\be
\bar Q = \frac{i}{2e^2}\int d^2x \Sigma^\dagger (1 + f + \gamma^0g)
(\gamma^0 B + D) + \frac{i}{2}\int d^2x \psi^\dagger \Dsl \phi
\label{silvaesmuycatolico}
\ee
with
\be
f = - 1 + \frac{B^2+D^2}{B^2-D^2} M - \frac{2BD}{(B^2-D^2)^2} N
\label{fs}
\ee
\be
g = \frac{B^2+D^2}{B^2-D^2}\left(
N - \frac{2BD}{(B^2-D^2)^2} M
\right)
\label{gs}
\ee
with
\be
M = 2\left(\frac{\beta^2 + B^2}{r} - \beta^2
\right)
\label{ms}
\ee
\be
N = \frac{2BD}{r}
\label{ns}
\ee
\be
r = \sqrt{1 + \frac{1}{\beta^2}(B^2 + D^2)}
\label{rs}
\ee
With this, we can compute the SUSY algebra
\be
\{ Q,\bar Q\} = \Psl + Z
\label{zzz}
\ee
and identify the energy and central charge in terms of $f$ and $g$,
\be
E = \frac{1}{2e^2} \int d^2x
\left( (B^2 + D^2)(f +1) +  2g BD
\right) + \frac{1}{2} \int d^2x |D_i\phi|^2
\label{ene}
\ee
\be
Z = \frac{1}{2e^2} \int d^2x
\left( (B^2 + D^2) g +  2(f+1) BD
\right) + \frac{1}{2} \int d^2x \varepsilon_{ij} D_i\phi^* D_j\phi
\label{zene}
\ee
Now, using the equation of motion for the auxiliary field $D$ (eq.(\ref{lab})
one can see that the r.h.s. in eqs.(\ref{ene})-(\ref{zene}) take the form
\be
E= \sqrt{
\left(
(1 + \frac{B^2}{\beta^2})
(1 + \frac{e^4}{4\beta^2}(\phi^2 -\xi^2)^2)
\right)} - \frac{\beta^2}{e^2} + \frac{1}{2}|D_i\phi|^2
\label{w}
\ee
\be
Z = \int d^2x
\left( B(\phi^2 - \xi^2) + \varepsilon_{ij}D_i\phi D_j\phi^*
\right) = \xi^2 \oint A_\mu dx^\mu = \xi^2 n
\label{u}
\ee
and then squaring (\ref{zzz}) one again gets the Bogomol'nyi
bound
\be
E \ge |Z|
\label{ez}
\ee
this showing the consistency of our supersymmetric construction. We stress
that the results summarized in eqs.(\ref{silvaesmuycatolico})-(\ref{ez})
correspond to the exact supersymmetric Born-Infeld model and not just
some approximation in powers of $1/\beta^2$.

\section{Summary and Discussion}
Studying the $N=2$ supersymmetric completion of the Born-Infeld-Higgs
model in $d=3$ dimensions, we have found the Bogomol'nyi relations
for the bosonic theory. The interest in $d=3$ dimensions arises from
the fact that in such space-time dimensions vortex solutions to
Bogomol'nyi equations are known to exist. Remarkably, we have found
that the same set of equations (and hence of solutions) hold when
a Born-Infeld Lagrangian determines the dynamics of the gauge
field.

Originally, supersymmetric extensions of the Born-Infeld theory
were constructed using the superfield formalism \cite{DP}-\cite{CF}
and only the bosonic sector was explicitly written in component fields.
Since one of our goals was to derive Bogomol'nyi relations from
the supersymmetry $N=2$ algebra, we needed the explicit form of
the fermionic Lagrangian, at least, up to quadratic terms in
the fermion fields which in turn lead to linear terms in the Noether current
which give the sole
non-vanishing contributions to the algebra in the
bosonic background sector.

Our analysis shows that  supersymmetry forces a particular functional form
of the bosonic action in which the Higgs potential enters in the
Born-Infeld square root (see eq.(\ref{ac}) in such a way as to ensure that
the same   Bogomol'nyi relations hold both for the Maxwell and
the Born-Infeld theory.

As it was to be expected, the central charge of the $N=2$ SUSY algebra
coincides with the topological charge (the number of
vortex magnetic flux units) of the model this ensuring that
the Bogomol'nyi bound is not modified when one has a Born-Infeld theory.
This was explicitly shown by constructing the SUSY algebra and
deriving from Bogomol'nyi inequality in the usual way.

{}~

\underline{Acknowledgments}:
We would like to thank A.~Lugo for helpful comments and
discussions. F.A.S. is
partially  supported by CICBA, CONICET and
Fundaci\'on Antorchas,   Argentina and a Commission of the European Communities
contract No:C11*-CT93-0315.

\newpage

\section*{Appendix: Superfields in components}

We start from the standard form for the chiral superfield $W_\alpha $ (see
for example ref. \cite{lykken})
\begin{equation}
W_\alpha \left( y,\theta ,\bar \theta \right) =-i\lambda _\alpha +\theta
_\alpha D-\frac i2\left( \sigma ^m\bar \sigma ^n\theta \right) _\alpha
F_{mn}+\theta \theta \left( \sigma ^m\partial _m\bar \lambda \right) _\alpha
\end{equation}
where $\lambda ,$ $\bar \lambda ,$ $D,$ $F$ and $\tilde F$ are functions of
the variable $y^m=x^m+i\theta \sigma ^m\bar \theta $ and $x^m$ is the usual
4-vector position (It will be convenient to write all the
superfields in terms of the variable $x$ instead of $y.$).
The covariant derivatives are defined in eqs.(\ref{10}) and
(\ref{11}).
\vspace{0.4 cm}

{\bf Components of }$W^\alpha W_\alpha ${\bf \ and }$\bar W_{\dot \alpha
}\bar W^{\dot \alpha }$
\vspace{0.05 cm}

\be
\left. W^2\left( x\right) \right| _0=-\lambda \lambda
\ee

\be
\left. W^2\left( x\right) \right| _\theta =-2i\theta \lambda D+\theta
\sigma ^\mu \bar \sigma ^\nu \lambda F_{\mu \nu }
\ee

\be
\left. W^2\left( x\right) \right| _{\theta \theta }=\theta \theta \left(
-2i\lambda \dsl \bar \lambda +A\right)
\ee

\be
\left. W^2\left( x\right) \right| _{\theta \bar \theta }=-i\theta \sigma
^\mu \bar \theta \partial _\mu \left( \lambda \lambda \right)
\ee
\be
\left. W^2\left( x\right) \right| _{\theta \theta \bar \theta }=-\theta
\theta \partial _\mu \left( \Omega ^{\mu \nu }\eta _{\nu \rho }\lambda
\sigma ^\rho \bar \theta \right)
\ee
\be
\left. W^2\left( x\right) \right| _{\theta \theta \bar \theta \bar \theta
}=\frac 14\theta \theta \bar \theta \bar \theta \Box \left( \lambda \lambda
\right)
\ee
\noindent with
\begin{eqnarray*}
A &=&D^2-\frac 12F^{\mu \nu }F_{\mu \nu }-\frac i2F^{\mu \nu }\tilde F_{\mu
\nu } \\
A^{*} &=&D^2-\frac 12F^{\mu \nu }F_{\mu \nu }+\frac i2F^{\mu \nu }\tilde
F_{\mu \nu }
\end{eqnarray*}
\be
\Omega ^{\mu\nu}=D\eta ^{\mu\nu}+iF^{\mu\nu}-\tilde F^{\mu\nu}
\ee
where $\sigma ^\mu $ and $\bar \sigma ^\nu $ are the Pauli matrices, defined
as in ref. \cite{lykken}. Everywhere, except if explicitly stated, $D,F_{\mu
\nu },$
$\tilde F_{\mu \nu }$ and $\lambda $ depend on $x.$

The components of $\bar W^2$ can be obtained from the former expressions by
calculating the adjoint of the matrix elements.
\vspace{0.4 cm}

{\bf Components of }$W^2\bar W^2(x)$
\vspace{0.05 cm}

\be
\left. W^2\bar W^2\left( x\right) \right| _0=\lambda \lambda \bar \lambda
\bar \lambda \
\ee

\be\left. W^2\bar W^2\left( x\right) \right| _\theta =2i\bar \lambda \bar
\lambda \left( \theta \lambda D-\frac i2\theta \sigma ^\mu \bar \sigma ^\nu
\lambda F_{\mu \nu }\right)
\ee

\be
\left. W^2\bar W^2\left( x\right) \right| _{\theta \theta }=-\theta \theta
\bar \lambda \bar \lambda \left[ -2i\lambda \dsl \bar \lambda
+A\right]
\ee

\begin{eqnarray}
\left. W^2\bar W^2\left( x\right) \right| _{\theta \bar \theta }& = &
-i\theta
\sigma ^m\bar \theta \left( \lambda \lambda \right) \stackrel{%
\longleftrightarrow }{\partial _m}\left( \bar \lambda \bar \lambda \right)
+4\left( \theta \lambda D-\frac i2\theta \sigma ^\mu \bar \sigma ^\nu
\lambda F_{\mu \nu }\right) \nonumber\\
& &  \left( \bar \theta \bar \lambda D+\frac i2\bar
\theta \bar \sigma ^\rho \sigma ^\sigma \lambda F_{\rho \sigma }\right)
\end{eqnarray}
\begin{eqnarray}
\left. W^2\bar W^2\left( x\right) \right| _{\theta \theta \bar \theta
} & = &
\theta \theta \left\{ \left( 2i\bar \theta \bar \lambda D-\bar \lambda
\bar \sigma ^\mu \sigma ^\nu \bar \theta F_{\mu \nu }\right) \left(
-2i\lambda \dsl \bar \lambda +A\right) +\right.
 \nonumber
\\
& & \left. \left( \bar \theta \bar
\sigma ^m\lambda \Omega ^{pq}\eta _{qm}\right) \stackrel{\longleftrightarrow
}{\partial _p}\left( \bar \lambda \bar \lambda \right) \right \}
\end{eqnarray}
with
\begin{eqnarray*}
\left.
W^2\bar W^2
\right|_{\theta \theta \bar \theta \bar \theta }
&=&\theta \theta \bar \theta \bar \theta
\left\{
-\frac 14
\left(
\lambda\lambda \Box \bar \lambda \bar \lambda
+\bar \lambda \bar \lambda \Box\lambda \lambda
\right)
+\frac 12
\partial _\mu
\left(
\lambda \lambda
\right)
\partial ^\mu
\left(
\bar \lambda \bar \lambda
\right)
\right.  \\
&&
-4
\left(
\lambda \dsl \bar \lambda
\right)
\left(
\bar \lambda \bar {\dsl}\lambda
\right)
\left.
-2iA^{*}\lambda \dsl \bar \lambda
-i\Omega ^{*\nu \rho
}\Omega _{\rho \mu }\lambda \sigma ^\mu \partial _\nu \bar \lambda
- \right.  \nonumber \\
&&
2iA\bar\lambda \bar \partial \lambda
-i\Omega ^{\nu \rho }\Omega _{\rho \mu
}^{*}\bar \lambda \bar \sigma ^\mu \partial _\nu \lambda
\nonumber\\
& & \left.
-i\partial _\nu
\left(
\Omega ^{*\nu \rho }
\right)
\Omega _{\rho
\mu }\lambda \sigma ^\mu \bar \lambda -i\partial _\nu
\left
( \Omega ^{\nu
\rho }
\right)
\Omega _{\rho \mu }^{*}\bar \lambda \bar \sigma ^\mu \lambda
+AA^{*}
\right
\}
\end{eqnarray*}
where
\[
A\!\stackrel{\leftrightarrow }{\partial }\!B=A\partial B-\left( \partial
A\right) B
\]
 \be
\Omega ^{*\nu \rho }\Omega _{\rho \mu }=
\left(
D^2+\frac 12F_{\alpha \beta
}F^{\alpha \beta }
\right)
\delta _\mu ^\nu -2D\eta ^{\nu \rho }\tilde
F_{\rho \mu }+2F^{\nu \rho }F_{\rho \mu }
\ee
and
\be
{Im}
\left(
\partial _\nu
\left(
\Omega ^{*\nu \rho }
\right)
\Omega
_{\rho \mu }
\right)
=-D\partial _\nu F^{\nu \mu }+\partial _\nu
\left(
F^{\nu \alpha }
\right)
\tilde F_{\alpha \mu }+\tilde F^{\nu \alpha }\partial
_\nu F_{\alpha \mu }+\frac 12\partial _\mu (F_{\alpha \beta })\tilde
F^{\alpha \beta }
\label{sonia}
\ee

(we will see later that we do not need the real part of this expression)

\vspace{0.4 cm}

{\bf Components of }$X\left( x\right) \equiv \frac 18\left( D^2W^2(x)+\bar
D^2\bar W^2(x)\right) $
\vspace{0.05 cm}

\be
\left. X\right| _0=\left( i\left( \lambda \dsl \bar \lambda +\bar
\lambda \bar {\dsl }\lambda \right) -D^2+\frac 12F^2\right)
\ee

\be
\left. X\right| _\theta =-\eta ^{qt}\left( \theta \sigma ^p\right)_{\dot
\alpha }\partial _q\left( \bar \lambda ^{\dot \alpha }\Omega
_{tp}^{*}\right)
\ee

\be
\left. X\right| _{\bar \theta }=-\eta ^{qt}\left( \bar \theta \bar \sigma
^p\right) ^\alpha \partial _q\left( \lambda _\alpha \Omega _{tp}\right)
\ee

\be
\left. X\right| _{\theta \bar \theta }=-i\theta \sigma ^p\bar \theta
\partial _p\left( i\left( \lambda \dsl \bar \lambda +\bar \lambda
\bar {\dsl }\lambda \right) -D^2+\frac 12F^2\right)
\ee

\be
\left. X\right| _{\theta \theta }=-\frac 12\theta \theta \Box \left( \bar
\lambda \bar \lambda \right)
\ee

\be
\left. X\right| _{\bar \theta \bar \theta }=-\frac 12\bar \theta \bar
\theta \Box \left( \lambda \lambda \right)
\ee

\be
\left. X\right| _{\theta \theta \bar \theta }=-\frac 12i\theta \theta
\left[ \Box \left( \theta ^\alpha a_{\alpha \beta }\lambda ^\beta \right)
-\partial _m\partial ^t\left( \theta \sigma ^m\bar \sigma ^n\lambda \Omega
_{tn}\right) \right]
\ee

\be
\left. X\right| _{\theta \bar \theta \bar \theta }=-\frac 12i\bar \theta
\bar \theta \left[ \Box \left( \theta ^\alpha a_{\alpha \beta }\lambda
^\beta \right) -\partial _m\partial ^t\left( \theta \sigma ^m\bar \sigma
^n\lambda \Omega _{tn}\right) \right]
\ee

\be
\left. X\right| _{\theta \theta \bar \theta \bar \theta }=-\frac 14\Box
\left( i\left( \lambda \dsl \bar \lambda +\bar \lambda \bar {
\dsl}\lambda \right) -D^2+\frac 12F^2\right)
\ee

\vspace{0.4 cm}

{\bf Components of }$Y\left( x\right) \equiv -\frac i{16}\left(
D^2W^2(x)-\bar D^2\bar W^2(x)\right) $
\vspace{0.05 cm}

\be
\left. Y\right| _0=\frac 12\left( \left( \lambda \dsl \bar \lambda
-\bar \lambda \bar {\dsl }\lambda \right) +\frac 12F\tilde F\right)
\ee

\be
\left. Y\right| _\theta =\frac i2\eta ^{qt}\left( \theta \sigma ^p\right)
_{\dot \alpha }\partial _q\left( \bar \lambda ^{\dot \alpha }\Omega
_{tp}^{*}\right)
\ee

\be
\left. Y\right| _{\bar \theta }=-\frac i2\eta ^{qt}\left( \bar \theta \bar
\sigma ^p\right) ^\alpha \partial _q\left( \lambda _\alpha \Omega
_{tp}\right)
\ee

\be
\left. Y\right| _{\theta \bar \theta }=-\frac i2\theta \sigma ^p\bar \theta
\partial _p\left( \left( \lambda \dsl \bar \lambda -\bar \lambda
\bar {\dsl }\lambda \right) +\frac 12F\tilde F\right)
\ee

\be
\left. Y\right| _{\theta \theta }=-\frac i4\theta \theta \Box \left( \bar
\lambda \bar \lambda \right)
\ee

\be
\left. Y\right| _{\bar \theta \bar \theta }=\frac i4\bar \theta \bar \theta
\Box \left( \lambda \lambda \right)
\ee

\be
\left. Y\right| _{\theta \theta \bar \theta }=-\frac 14\bar \theta \bar
\theta \left[ \Box \left( \theta ^\alpha a_{\alpha \beta }\lambda ^\beta
\right) -\partial _m\partial ^t\left( \theta \sigma ^m\bar \sigma ^n\lambda
\Omega _{tn}\right) \right]
\ee

\be
\left. Y\right| _{\theta \theta \bar \theta \bar \theta }=-\frac 18\Box
\left( \left( \lambda \dsl \bar \lambda -\bar \lambda \bar
{\dsl }\lambda \right) +\frac 12F\tilde F\right)
\ee

\end{document}